\documentclass[aps,prb,twocolumn,letterpaper,superscriptaddress]{revtex4}

\usepackage{times}
\usepackage[T1]{fontenc}
\usepackage{color}
\usepackage{graphicx}
\usepackage{mathtools,amsmath,amssymb,amsfonts,bm}
\usepackage{cases}

\usepackage{units}
\usepackage[overload]{empheq}

\usepackage{ulem}
\usepackage{multirow}
\usepackage{dcolumn}
\usepackage{tabularx}
\newcolumntype{Y}{>{\centering\arraybackslash}X}
\newcolumntype{Z}{>{\arraybackslash}X}

\graphicspath{{./}}

\renewcommand{\cite}[1]{[\onlinecite{#1}]}

\newcommand{\bi}[1]{\bm{\mathit{#1}}}
\newcommand{\gowo}{G$_0$W$_{0}$}

\begin{document}

\title{Combined first-principles and model Hamiltonian study\\of the
  perovskite series $R$MnO$_{3}$ ($R$ = La, Pr, Nd, Sm, Eu and Gd)}
\date{\today}

\author{Roman Kov\'a\v{c}ik}

\email{r.kovacik@fz-juelich.de}

\affiliation{\mbox{Peter Gr\"unberg Institut and Institute for Advanced
  Simulation, Forschungszentrum J\"ulich and JARA, 52425 J\"ulich,
  Germany}}

\author{Sowmya Sathyanarayana Murthy}

\affiliation{University of Vienna, Faculty of Physics, Sensengasse
  8/8, A-1090 Wien, Austria}

\author{Carmen E. Quiroga}

\affiliation{Dept. of Earth and Environmental Sciences,
  Ludwig-Maximillians-Universit\"at M\"unchen, Theresienstrasse 41,
  80333 Munich, Germany}

\author{Claude Ederer}

\affiliation{Materials Theory, ETH Z\"urich, Wolfgang-Pauli-Strasse
  27, 8093 Z\"urich, Switzerland}

\author{Cesare Franchini}

\affiliation{University of Vienna, Faculty of Physics, Sensengasse
  8/8, A-1090 Wien, Austria}

\begin{abstract}
  We merge advanced \emph{ab initio} schemes (standard density
  functional theory, hybrid functionals and the GW approximation) with
  model Hamiltonian approaches (tight-binding and Heisenberg
  Hamiltonian) to study the evolution of the electronic, magnetic and
  dielectric properties of the manganite family $R$MnO$_{3}$ ($R$ =
  La, Pr, Nd, Sm, Eu and Gd).  The link between first principles and
  tight-binding is established by downfolding the physically relevant
  subset of $3d$ bands with $e_g$ character by means of maximally
  localized Wannier functions (MLWFs) using the VASP2WANNIER90
  interface. The MLWFs are then used to construct a general
  tight-binding Hamiltonian written as a sum of the kinetic term, the
  Hund's rule coupling, the JT coupling, and the electron-electron
  interaction. The dispersion of the TB $e_g$ bands at all levels are
  found to match closely the MLWFs. We provide a complete set of TB
  parameters which can serve as guidance for the interpretation of
  future studies based on many-body Hamiltonian approaches.  In
  particular, we find that the Hund's rule coupling strength, the
  Jahn-Teller coupling strength, and the Hubbard interaction parameter
  $U$ remain nearly constant for all the members of the $R$MnO$_{3}$
  series, whereas the nearest neighbor hopping amplitudes show a
  monotonic attenuation as expected from the trend of the tolerance
  factor.  Magnetic exchange interactions, computed by mapping a large
  set of hybrid functional total energies onto an Heisenberg
  Hamiltonian, clarify the origin of the A-type magnetic ordering
  observed in the early rare-earth manganite series as arising from a
  net negative out-of-plane interaction energy. The obtained exchange
  parameters are used to estimate the N\'eel temperature by means of
  Monte Carlo simulations. The resulting data capture well the
  monotonic decrease of the ordering temperature down the series from
  $R$ = La to Gd, in agreement with experiments. This trend correlates
  well with the modulation of structural properties, in particular
  with the progressive reduction of the Mn-O-Mn bond angle which is
  associated with the quenching of the volume and the decrease of the
  tolerance factor due to the shrinkage of the ionic radii of $R$
  going from La to Gd.
\end{abstract}

\maketitle

\section{Introduction}\label{sec:intro}

Perovskite transition metal oxides (TMOs), which fall under the
category of strongly correlated systems, exhibit a wide array of
complex orbitally and spin ordered states, arising from the interplay
of the structural, electronic and magnetic degrees of freedom. In
particular, rare earth manganites with the general formula
$R_{1-x}A_{x}$MnO$_{3}$, where $R$ is a trivalent rare earth cation
and $A$ is a divalent alkaline earth cation, exhibit stunning
characteristics such as the colossal magnetoresistance (CMR)
effect~\cite{1993-10//von-helmolt/wecker//samwer,2001-08//salamon/jaime,1997-04//khomskii/sawatzky,1989-03//kusters/singleton//hayes,1994-04//jin/tiefel//chen},
observed in compounds like Pr$_{1-x}$Ca$_{x}$MnO$_{3}$,
Pr$_{1-x}$Ba$_{x}$MnO$_{3}$, Nd$_{0.5}$Sr$_{0.5}$MnO$_{3}$ and in the
well-known hole-doped
LaMnO$_{3}$~\cite{1994-11//tokura/urushibara//furukawa,1951-05//zener}.
Another interesting property, tuned by the Mn$^{3+}$ magnetic
structure variation in
$R$MnO$_{3}$~\cite{2003-08//kimura/ishihara//tokura}, is the emergence
of magneto-electric/multiferroic properties for the smaller rare earth
cations ($R=$ Gd, Tb,
Dy)~\cite{2004-06//goto/kimura//tokura,2005-06//kimura/lawes//ramirez}.
Despite the large number of studies on CMR and parent CMR compounds,
experimental~\cite{2003-01//sanchez/subias//blasco,2009-02//ferreira/agostinho-moreira//mendonca,2009-03//chatterji/schneider//bhattacharya,2000-02//munoz/alonso//fernandez-diaz,1983-03//kamegashira/miyazaki,2005-03//dabrowski/kolesnik//mais,2006-06//laverdiere/jandl//iliev,2006-02//iliev/abrashev//sun,2009-05//lee/kida//tokura,2012-01//iyama/jung//kimura}
and theoretical
studies~\cite{2008-07//yamauchi/freimuth//picozzi,2009-08//kim/min,2010-07//moskvin/makhnev//balbashov,2009-10//dong/yu//dagotto,2009-07//choithrani/rao//singh,2012-12//he/franchini,2012-05//he/chen//franchini,2012-06//franchini/kovacik//kresse,2014-06//franchini}
on early $R$MnO$_{3}$ are found in less numbers.

\begin{figure}[b!]
  \centering
  \includegraphics{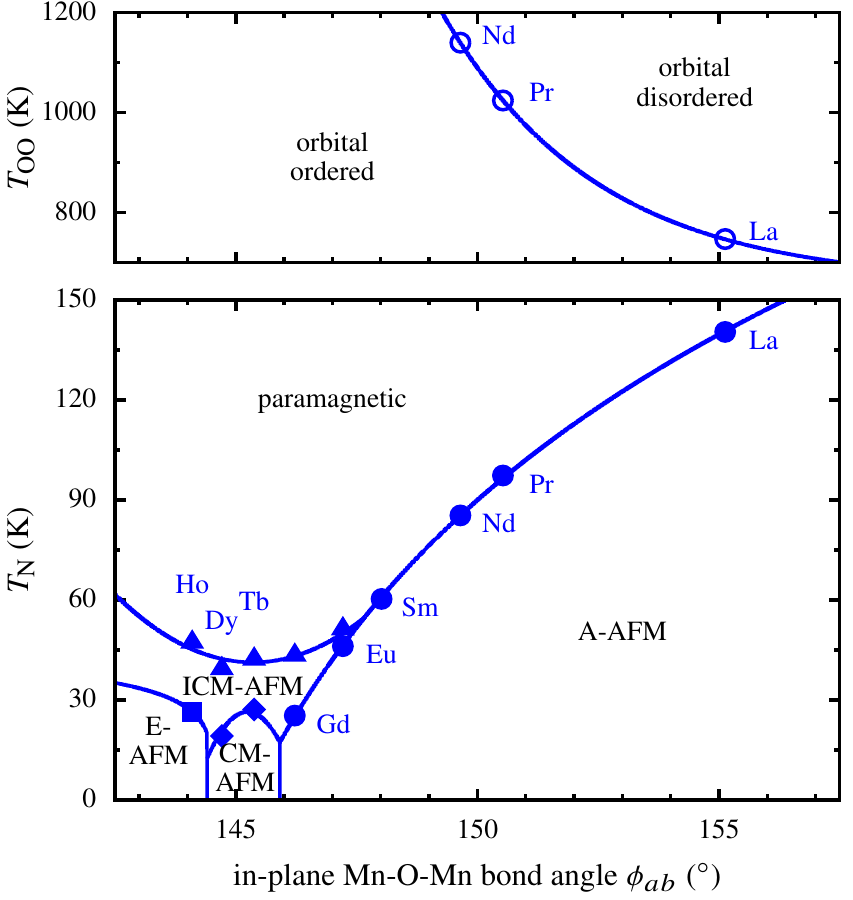}
  \caption{Phase diagram of the orbital (top) and spin (bottom) order
    in the early series of $R$MnO$_{3}$ as a function of the in-plane
    Mn-O-Mn bond angle $\phi_{ab}$. Adapted from
    Ref.~\cite{2003-08//kimura/ishihara//tokura}}
  \label{fig:RMnO3phase}
\end{figure}

The phase diagram of $R$MnO$_{3}$ (Fig.\ref{fig:RMnO3phase}) reported
by Kimura~\emph{et al.}~\cite{2003-08//kimura/ishihara//tokura} shows
the trends of the orbital ($T_\mathrm{OO}$) and the spin
($T_\mathrm{N}$) ordering temperatures as a function of the in-plane
Mn-O-Mn angle $\phi_{ab}$. It also illustrates that when the La$^{3+}$
cation is replaced by smaller cations, a successive increase in the
orthorhombic distortion, manifested by the decrease of $\phi_{ab}$, is
observed. The orbital ordering temperature $T_\mathrm{OO}$
monotonically increases with the decreasing atomic radius $r_{R}$ of
cation $R$, whereas the spin-ordering temperature $T_\mathrm{N}$
decreases steadily from 140~K for LaMnO$_{3}$ to 40~K for GdMnO$_{3}$
with decreasing $r_{R}$. The Mn-O-Mn bond angle is reduced by the
smaller $R^{3+}$ ion at the $A$ site, which in turn increases the
tilting of the oxygen octahedra, thereby weakening the A-type
anti-ferromagnetic (A-AFM) order, characterized by an in-plane
parallel alignment of spins antiparallelly coupled to the spins in
adjacent planes.

Understanding the microscopic details of the manganite systems could
help to gain insights into the fundamental physics behind these
interesting phenomena. Theoretically, TMOs have been historically
studied using two different approaches: \emph{ab initio} and model
Hamiltonians typically based on a tight-binding parameterization.
With regard to first-principles calculations on $R$MnO$_{3}$,
particularly detailed and interesting theoretical findings have been
reported by Yamauchi~\emph{et
  al.}~\cite{2008-07//yamauchi/freimuth//picozzi}, where the authors
discuss the validity of the commonly used generalized gradient
approximation (GGA) to the exchange-correlation (XC) functional within
the density functional theory (DFT) for $R$MnO$_{3}$ compounds. By
adopting the fully optimized structure, it was shown that the
Jahn-Teller (JT) distortion, typical of manganite systems and
manifested by an alternating Mn-O bond length disproportionation, is
underestimated using GGA.  In agreement with the earlier study of
Yin~\emph{et al.}~\cite{2006-03//yin/volja/ku}, the situation in
LaMnO$_{3}$ improves by incorporating an on site Hubbard $U$ parameter
to the GGA or to the local density approximation (LDA), while for the
other compounds in the series the agreement with the experimental
structural data worsens. Similarly, the orthorhombic distortion in the
whole series is better captured using the GGA approach. Finally, for
values of ${U\geq{4}}$~eV, the ferromagnetic (FM) ordering becomes the
most favorable contrary to the experimental observation of A-AFM
ordering. However, the deficiency of GGA in predicting the magnetic
properties was also pointed out. While experiments have shown that at
\mbox{$T=0$~K}, the A-AFM state is the spin ground state even in
GdMnO$_{3}$, GGA shows a total energy trend where the E-type AFM
(E-AFM) and the A-AFM phases are degenerate in SmMnO$_{3}$ and the
E-AFM phase is found to be the most stable ordering for GdMnO$_{3}$.

In this study, we aim to investigate the evolution of the electronic
and magnetic properties in the early series of $R$MnO$_{3}$ ($R$ = La,
Pr, Nd, Sm, Eu, Gd). By combining first-principles calculations and
the tight binding (TB) approach via maximally localized Wannier
functions (MLWFs), we calculate the TB parameters by applying the
methodology that was described in
Ref.~\cite{2012-06//franchini/kovacik//kresse} for LaMnO$_{3}$. Two
alternative model parameterizations are considered, which account for
the effects of the electron-electron (el-el) interaction either
implicitly in the otherwise non-interacting TB parameters or
explicitly via a mean-field el-el interaction term in the TB
Hamiltonian. Using this methodology, we explore the changes in the
band structure of $R$MnO$_{3}$ and construct, compare and interpret
the obtained TB parameters. Different levels of approximation to the
XC kernel are adopted: standard DFT within GGA, hybrid functionals,
and GW. Thereby a ready-to-use set of TB parameters is provided for
future studies.

We will start with a brief overview of the basic ground state
properties of the $R$MnO$_{3}$ series (Sec.~\ref{sec:struc}) followed
by two methodological sections focused on the description of the
tight-binding parametrization (Sec.~\ref{sec:tb}) and the {\em ab
  initio} calculations (Sec.~\ref{sec:ab}).  The results for the
electronic structure, magnetic properties and tight binding parameters
are presented and discussed in Sec.~\ref{sec:res}. The article ends
with a brief summary and conclusions.

\section{The $R$MnO$_{3}$ series: fundamentals}\label{sec:struc}

\begin{figure}[b!]
  \centering
  \includegraphics{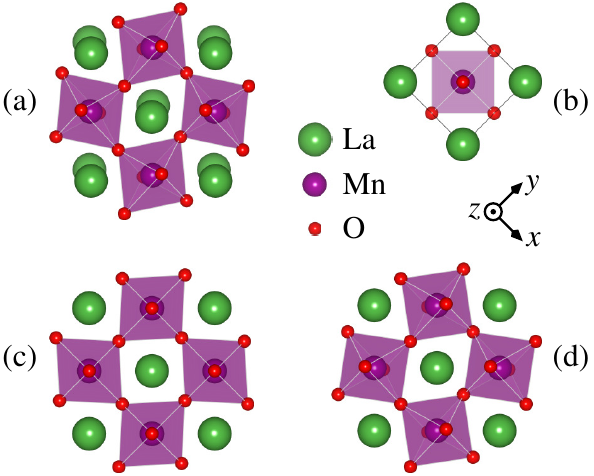}
  \caption{(a) The experimental $R$MnO$_{3}$ crystal structure
    (example of $R=\mathrm{La}$) with distortion modes imposed on (b)
    the simple cubic perovskite structure: (c) the pure JT $Q^x$ mode
    distortion and (d) the pure GFO-type distortion. Structural models
    were generated with VESTA~\cite{2011-12//momma/izumi}. Adapted
    from Fig.~1 in Ref.~\cite{2010-06//kovacik/ederer}.}
  \label{fig:rmo-structures}
\end{figure}

The ground state electronic structure of $R$MnO$_{3}$ ($R=$ La, Pr,
Nd, Sm, Eu and Gd) is characterized by the crystal-field induced
breaking of the degeneracy of the Mn$^{3+}$ $3d^{4}$ manifold in the
high-spin configuration ($t_{2g}$)$^{3}$ ($e_{g}$)$^{1}$, with the
$t_{2g}$ orbitals lying lower in energy than the two-fold degenerate
$e_{g}$ ones. Due to the strong Hund's rule coupling, the spins in the
fully occupied majority $t_{2g}$ orbitals ${(S=3/2)}$ are aligned
parallel to the spin in the singly occupied majority $e_{g}$ state
${(S=1/2)}$ at the same site. The orbital degeneracy in the $e_{g}$
channel is further lifted via cooperative Jahn-Teller
distortions~\cite{1998-02//rodriguez-carvajal/hennion//revcolevschi,2003-08//chatterji/fauth//ghosh,2003-01//sanchez/subias//blasco,2005-05//qiu/proffen//billinge},
manifested by long and short Mn-O octahedral bonds alternating within
the conventional orthorhombic basal plane, which are accompanied by
GdFeO$_{3}$-type (GFO) checkerboard tilting and rotations of oxygen
octahedra~\cite{1971-05//elemans/van-laar//loopstra,1995-10//norby/krogh-andersen//andersen,1997-02//woodward}
(see Fig.~\ref{fig:rmo-structures}).
As a result, the ideal cubic
perovskite structure is strongly distorted into an orthorhombic
structure with \textit{Pbnm}
symmetry~\cite{1971-05//elemans/van-laar//loopstra,1995-10//norby/krogh-andersen//andersen}
and it has been experimentally confirmed that the orbital ordering is
of C-type, where the occupied $e_g$ orbitals follow the checkerboard
JT distortion pattern in the $xy$-plane and the planes are stacked
along the $z$-axis~\cite{1998-07//murakami/hill//endoh}. The occupied
$e_{g}$ orbital can be represented by a linear combination of the
$d_{z^2}$ and $d_{x^2-y^2}$ character orbitals as
${|\theta\rangle =\mathrm{cos}\frac{\theta}{2}|3z^2-r^2\rangle
  +\mathrm{sin}\frac{\theta}{2}|x^2 -y^2\rangle}$,
where $\theta$ is the orbital mixing
angle~\cite{2006-03//yin/volja/ku,1960-05//kanamori,2010-02//pavarini/koch,2003-02//sikora/oles}.

\begin{figure}[b!]
\centering
\includegraphics{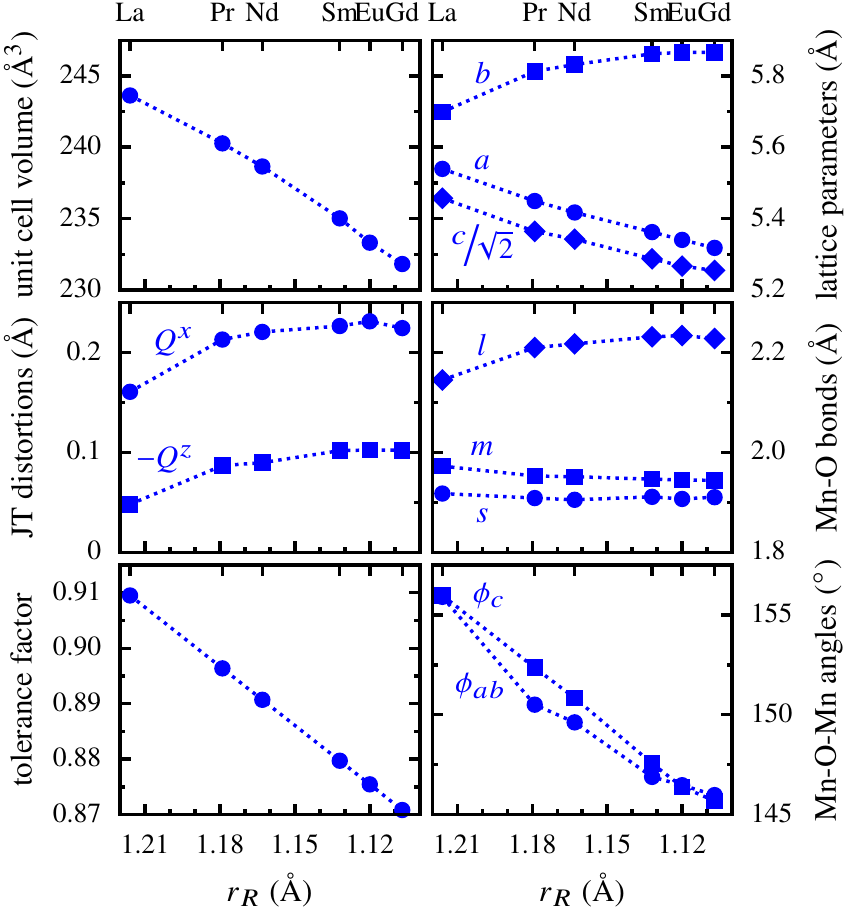}
\caption{(color online) Dependence of various structural parameters on
  the $R$ cation radius $r_R$, taken or calculated from the
  experimental crystal structures at room temperature. LaMnO$_{3}$
  data are taken from
  Ref.~\cite{1995-10//norby/krogh-andersen//andersen}, $R$MnO$_{3}$
  ($R=$ Pr, Nd) from
  Ref.~\cite{2000-03//alonso/martinez-lope//fernandez-diaz} and
  $R$MnO$_{3}$ \mbox{($R=$ Sm, Eu, Gd)} from
  Ref.~\cite{2002-05//mori/kamegashira//fukuda}. The tolerance factor
  is calculated using the ionic radii listed in
  Ref.~\cite{1976-09//shannon}. }
\label{fig:struc-RMnO3}
\end{figure}

The most important structural characteristics as a function of the
rare earth cation radius $r_R$ are collected in
Fig.~\ref{fig:struc-RMnO3}. 
As $r_{R}$ decreases from La to Gd, the major effect is the unit cell
volume $V$ reduction associated with the progressive decrease of the
lattice parameters $a$ and $c$ (the so called ``lanthanide
contraction''). In
Ref.~\cite{2000-03//alonso/martinez-lope//fernandez-diaz}, it was
pointed out that the characteristic relation ${c/\sqrt{2}<a<b}$ has
its origin in the strong cooperative JT effect, inducing orbital
ordering and distorting the MnO$_{6}$ octahedra. The local JT
distortion modes are defined as ${Q^{x}=(l-s)/\sqrt{2}}$ and
${Q^{z}=(2m-l-s)/\sqrt{6}}$, where $l$, $s$ and $m$ stand for long,
short and medium Mn-O bond lengths, respectively. From the trend shown
in Fig.~\ref{fig:struc-RMnO3}, a sizable increase of both
${\lvert{Q^{z}}\rvert}$ and $Q^{x}$ can be seen in all members of the
series as compared to LaMnO$_3$, stemming from the increase in $l$
while $m$ and $s$ remain almost unchanged.

Another important quantity in the physics of $AB$O$_3$ compounds is
the tolerance factor $t$~\cite{1926-05//goldschmidt}, which gives an
indication on the degree of structural distortions and the stability
of the perovskite crystal structure. It can be defined as
${t=(r_{A}+r_\mathrm{O})/\big[\sqrt{2}(r_{B}+r_\mathrm{O})\big]}$,
where $r_{A}$, $r_{B}$ and $r_\mathrm{O}$ are the ionic radii of $A$,
$B$ and O, respectively. For the simple cubic perovskite structure,
${t=1}$. Depending on the magnitude of $t$, different crystal
structures are formed. In $R$MnO$_3$, the \mbox{$A=R$} cations are too
small to completely fill the space in the cubic structure. In this
situation, the MnO$_6$ octahedra undergo collective rotations to
maximize the space filling, thereby reducing the Mn-O-Mn bond angles
from the ideal $180^\circ$.  Clearly, the trend of the tolerance
factor $t$ is in accordance with the trend of the Mn-O-Mn bond
angles. According to Zhou and
Goodenough~\cite{2003-08//zhou/goodenough}, the transition temperature
$T_\mathrm{N}$ depends linearly on $\langle\cos^{2}\phi\rangle$, where
the average is taken over the three distinguishable Mn-O-Mn bond
angles, i.e. the two $\phi_{ab}$ bond angles in the $ab$-plane and the
$\phi_{c}$ bond angle in the $c$-direction.

\section{Methodology: Tight binding parameterization}\label{sec:tb}

Within the TB formalism, the effective electronic Hamiltonian of the
$e_{g}$ character manifold in manganites is generally written as a sum
of the following contributions: the kinetic energy term
$\hat{H}_\mathrm{kin}$ and several local interaction terms such as the
Hund's rule coupling to the $t_{2g}$ core spin
$\hat{H}_\mathrm{Hund}$, the JT coupling to the oxygen octahedra
distortion $\hat{H}_\mathrm{JT}$ and the electron-electron interaction
$\hat{H}_\mathrm{el-el}$~\cite{2007-10//ederer/lin/millis,2011-08//kovacik/ederer,2010-06//kovacik/ederer,2012-06//franchini/kovacik//kresse}:
\begin{align}
  \hat{H}_{\mathrm{kin}}=&
  -\sum_{\substack{\bi{R},\Delta\bi{R},\sigma\\a,b}}
  \hat{c}^{\sigma\dagger}_{a\left(\bi{R}+\Delta\bi{R}\right)}
  t^{\sigma}_{a\left(\bi{R}+\Delta\bi{R}\right)b\left(\bi{R}\right)}
  \hat{c}^{\sigma}_{b\left(\bi{R}\right)}\,,
  \label{eq:tb-kin}\\
  \hat{H}_{\mathrm{Hund}}=&
  -J_{\mathrm{H}}\sum_{\substack{\bi{R},\sigma,\sigma'\\a}}
  {\bi{S_R}}
  \hat{c}^{\sigma\dagger}_{a\left(\bi{R}\right)}
  {\boldsymbol{\tau}_{\sigma\sigma'}}
  \hat{c}^{\sigma'}_{a\left(\bi{R}\right)}\,,
  \label{eq:tb-hund}\\
  \hat{H}_{\mathrm{JT}}=&
  -\lambda\sum_{\substack{\bi{R},\sigma,i\\a,b}}
  \hat{c}^{\sigma\dagger}_{a\left(\bi{R}\right)}
  Q^{i}_{\bi{R}}\tau_{ab}^{i}
  \hat{c}^{\sigma}_{b\left(\bi{R}\right)}\,,
  \label{eq:tb-jt}\\
  \hat{H}_{\mathrm{el-el}}=&
  \frac{1}{2}\sum_{\substack{\bi{R},\sigma,\sigma'\\a,b,c,d}}
  U_{abcd}
  \hat{c}^{\sigma\dagger}_{a(\bi{R})}
  \hat{c}^{\sigma'\dagger}_{b(\bi{R})}
  \hat{c}^{\sigma'}_{d(\bi{R})}
  \hat{c}^{\sigma}_{c(\bi{R})}\,.
  \label{eq:tb-ee}
\end{align}
The annihilation $\hat{c}^{\sigma}_{a\left(\bi{R}\right)}$ and the
creation $\hat{c}^{\sigma\dagger}_{a\left(\bi{R}\right)}$ operators
are associated with orbital $\vert{a}\rangle$ at a particular Mn site
$\bi{R}$ (not to be confused with cation $R$) and spin $\sigma$. In
the kinetic energy term,
$t^{\sigma}_{a\left(\bi{R}+\Delta\bi{R}\right)b\left(\bi{R}\right)}$
is the hopping parameter between orbital $\vert{b}\rangle$ at site
$\bi{R}$ and orbital $\vert{a}\rangle$ at site
$\bi{R}+\Delta\bi{R}$. Further on, $J_{\mathrm{H}}$ is the Hund's rule
strength of coupling to the normalized $t_{2g}$ core spin $\bi{S_R}$,
$\lambda$ is the JT coupling constant and $Q^{i}_{\bi{R}}$ is the
amplitude of the particular JT mode $(i=\{x,z\})$ and $\tau_{ab}^{i}$
are the standard Pauli matrices. In this study, the electron-electron
interaction term is treated within a mean-field approximation
following the approach of Dudarev \emph{et
  al.}~\cite{1998-01//dudarev/botton//sutton}, involving a single
parameter ${U_\mathrm{W}=U_{aaaa}=U_{abab}}$, with all other
interaction matrix elements set to zero.

To obtain the model parameters we have extended the work presented in
Ref.~\cite{2012-06//franchini/kovacik//kresse} to the $R$MnO$_{3}$
early series, wherein the model parameters are obtained from the
Hamiltonian matrix elements in the MLWF basis. We will use a
simplified notation for the MLWF matrix elements with the two basis
functions of ${\lvert 3z^2-r^2 \rangle}$ and
${\lvert x^2-y^2 \rangle}$ character centered at the same
site. Thereby, the MLWF matrix element $h^{\Delta\bi{T}}_{mn}$, where
$\Delta\bi{T}$ is the lattice translation and $m$ and $n$ are general
orbital-site indices, can be written as:
$h^{\Delta\bi{T}}_{mn} \rightarrow h^{\Delta\bi{T}}_{a\bi{R},b\bi{R}'}
\rightarrow h^{\Delta\bi{R}}_{ab}$,
where $\Delta\bi{R}=\bi{R}'-\bi{R}+\Delta\bi{T}$. In order to
disentangle the effect of the JT distortion from other lattice
distortions, the TB model parameters are obtained from two crystal
structures: the experimental and the purely JT $Q^x$ mode distorted
structure, defined by the projection of the differences in the Wyckoff
positions of the experimental and the simple cubic perovskite
structure to the JT $Q^x$ mode (see Table~\ref{tab:wyckoff}). We note
that in this study we use the room temperature crystal
structures~\cite{1995-10//norby/krogh-andersen//andersen,2000-03//alonso/martinez-lope//fernandez-diaz,2002-05//mori/kamegashira//fukuda}
to maintain a consistent reference for all members of the $R$ series
given the available experimental data. Therefore, the results for
LaMnO$_3$ will differ from those in
Ref.~\cite{2012-06//franchini/kovacik//kresse}, where the low
temperature (4.2~K) structure from
Ref.~\cite{1971-05//elemans/van-laar//loopstra} was used in turn.

\begin{table}
  \centering
  \caption{Wyckoff positions of the $R=$ La, Pr, Nd, Sm, Eu, Gd site
    $(4c)$ and the two inequivalent oxygen sites O1 $(4c)$ and O2 $(8d)$ in the room
    temperature \textit{Pbnm} experimental structures (Expt.), taken from Refs.~\cite{1995-10//norby/krogh-andersen//andersen,2000-03//alonso/martinez-lope//fernandez-diaz,2002-05//mori/kamegashira//fukuda} and described in
    Fig.~\ref{fig:struc-RMnO3}, and their decomposition into
    the unit cell volume preserving structures with only the $Q^x$
    distortion mode (JT). Mn cations are at the high symmetry site
    $(4b)$ with Wyckoff positions ($\nicefrac{1}{2}$, 0, 0).}
  \smallskip
    \begin{tabularx}{246pt}{llYY}
      \hline\hline
      & &\multicolumn{2}{c}{Wyckoff Positions} \\
      \cline{3-4}
      &  & Expt. & JT\\
      \hline
 \multirow{3}{*} {LaMnO$_{3}$} & La    & (0.9937, 0.0435, \nicefrac{1}{4}) & (0.0, 0.0, \nicefrac{1}{4})\\
                              &    O1 & (0.0733, 0.4893, \nicefrac{1}{4}) & (0.0, 0.5, \nicefrac{1}{4})\\
                              &    O2 & (0.7257, 0.3014, 0.0385) & (0.7635, 0.2636, 0.0)\\
\multirow{3}{*} {PrMnO$_{3}$} & Pr    & (0.9911, 0.0639, \nicefrac{1}{4}) & (0.0, 0.0, \nicefrac{1}{4})\\
                              &    O1 & (0.0834, 0.4819, \nicefrac{1}{4}) & (0.0, 0.5, \nicefrac{1}{4})\\
                              &    O2 & (0.7151, 0.3174, 0.0430) & (0.7662, 0.2662, 0.0)\\
\multirow{3}{*} {NdMnO$_{3}$} & Nd    & (0.9886, 0.0669, \nicefrac{1}{4}) & (0.0, 0.0, \nicefrac{1}{4})\\
                              &    O1 & (0.0878, 0.4790, \nicefrac{1}{4}) & (0.0, 0.5, \nicefrac{1}{4})\\
                              &    O2 & (0.7141, 0.3188, 0.0450) & (0.7664, 0.2665, 0.0)\\
\multirow{3}{*} {SmMnO$_{3}$} & Sm    & (0.9850, 0.0759, \nicefrac{1}{4}) & (0.0, 0.0, \nicefrac{1}{4})\\
                              &    O1 & (0.0970, 0.4730, \nicefrac{1}{4}) & (0.0, 0.5, \nicefrac{1}{4})\\
                              &    O2 & (0.7076, 0.3241, 0.0485) & (0.7659, 0.2658, 0.0)\\
\multirow{3}{*} {EuMnO$_{3}$} & Eu    & (0.9841, 0.0759, \nicefrac{1}{4}) & (0.0, 0.0, \nicefrac{1}{4})\\
                              &    O1 & (0.1000, 0.4700, \nicefrac{1}{4}) & (0.0, 0.5, \nicefrac{1}{4})\\
                              &    O2 & (0.7065, 0.3254, 0.0487) & (0.7660, 0.2660, 0.0)\\
\multirow{3}{*} {GdMnO$_{3}$} & Gd    & (0.9384, 0.0807, \nicefrac{1}{4}) & (0.0, 0.0, \nicefrac{1}{4})\\
                              &    O1 & (0.1030, 0.4710, \nicefrac{1}{4}) & (0.0, 0.5, \nicefrac{1}{4})\\
                              &    O2 & (0.7057, 0.3246, 0.0508) & (0.7651, 0.2651, 0.0)\\
      \hline\hline
    \end{tabularx}
  \label{tab:wyckoff}
\end{table}

Two types of model parameterizations are employed, namely, Model~1 and
Model~2. Model~1 is an effectively "non-interacting" case, in which
the $\hat{H}_\mathrm{el-el}$ term is neglected with the purpose of
exploring how the more sophisticated beyond-PBE treatment of the XC
kernel affects the hopping, JT- and GFO-distortion related
parameters. Model~2 is an alternative way, involving an explicit
treatment of $\hat{H}_\mathrm{el-el}$ in the model Hamiltonian within
the mean-field approximation. This allows to obtain estimates of the
corresponding on-site interaction parameter by keeping the PBE on-site
model parameters as reference (see below).

In the following, for completeness, the considered TB model parameters
are briefly described. For more details on the practical use of the
VASP2WANNIER90 interface, as well as the derivation of the model
parameters used in this study, we refer to
Ref.~\cite{2012-06//franchini/kovacik//kresse}.

\subsection{Hopping parameters}

The kinetic energy is parameterized with seven parameters: four
hopping amplitudes and the JT distortion induced splitting in the
nearest neighbor hopping matrix, all evaluated in the purely JT $Q^x$
mode distorted structure, and two spin-dependent reduction parameters
of the hoppings due to the GFO distortion. For notation clarity, we
set the origin ($\bi{R}=\mathbf{0}$) at one of the Mn sites and align
the $x$ and $y$ cartesian axes with the direction of the long and
short Mn-O bonds of the JT $Q^x$ mode, respectively. The vectors
$\hat{\bi{x}}$, $\hat{\bi{y}}$, $\hat{\bi{z}}$ correspond to the
nearest-neighbor spacing of the Mn sites along the respective
axes~\cite{2012-06//franchini/kovacik//kresse}.

Nearest-neighbor hopping amplitudes $t^{ss}$ between sites within the
FM planes ($t^{\uparrow\uparrow}$, $t^{\downarrow\downarrow}$, $s$ is
a local spin index) are obtained as
${t^{ss}=\big({h_{11}^x}-3{h_{22}^x}\big)/2}$. The hopping parameter
between sites with antiparallel spin alignment is calculated as
${t^{ss'}=\big(t^{\uparrow\uparrow}+t^{\downarrow\downarrow}\big)/2}$.
The corresponding hopping matrices are then expressed as
${{\bf{t}}^{ss'}(\pm\hat{\bi{z}})=
  -t^{ss'}\big({\bf{I}}+{\boldsymbol{\tau}}^{z}\big)/2}$
and
${{\bf{t}}^{ss}(\pm\hat{\bi{x}})= -t^{ss}
  \big(2\,{\bf{I}}-\sqrt{3}\,{\boldsymbol{\tau}}^{x}-{\boldsymbol{\tau}}^{z}\big)/4}$,
where $\bf{I}$ is the unity matrix. Here and in the following, the
matrices along $\pm\hat{\bi{y}}$ are simply obtained by the relevant
symmetry transformation of the matrices along $\pm\hat{\bi{x}}$.

The JT distortion induces a splitting between the non-diagonal
elements of the nearest-neighbor hopping matrix. We model it as
${\Delta{\bf{t}}^{ss}(\pm\hat{\bi{x}})=\mathrm{i}\,\widetilde{\lambda}\,Q^x_{\bi{R}}
  {\boldsymbol{\tau}}^{y}}$,
where the parameter $\widetilde{\lambda}$ is obtained as
${\widetilde{\lambda}=\sum_{s}{\big(h_{12}^{x}-h_{21}^{x}\big)^{s}}/\left(4Q^{x}\right)}$.

The second-nearest neighbor hopping $t^{xy}$ and the second-nearest
neighbor hopping along the $x$, $y$, $z$ crystal axes $t^{2z}$ are
obtained as ${t^{xy}=-\sum_s\big(h_{11}^{xy}\big)^s/2}$ and
${t^{2z}=-\sum_s\big(h_{11}^{2z}\big)^s/2}$. While the hopping
matrices related to $t^{2z}$ have the same form as those of $t^{ss}$,
the second-nearest neighbor hopping matrices are expressed via
${{\bf{t}}(\pm\hat{\bi{x}}\pm\hat{\bi{z}})=
  t^{xy}\big({\bf{I}}-\sqrt{3}\,{\boldsymbol{\tau}}^{x}+{\boldsymbol{\tau}}^{z}\big)}$
and
${{\bf{t}}(\pm\hat{\bi{x}}\pm\hat{\bi{y}})=t^{xy}\big({\bf{I}}-2\,{\boldsymbol{\tau}}^{z}\big)}$.

In GFO distorted structures, all hopping matrices are scaled by a
spin-dependent reduction factor $(1-\eta_t^s)$, where
\mbox{$\eta_{t}^{s}=1-t^{ss}_{(\textit{Pbnm})}/t^{ss}$}. The hopping
parameter $t^{ss}_{(\textit{Pbnm})}$ is obtained analogously to the
$t^{ss}$ defined above but in the experimental crystal structure.

\subsection{On-site parameters}

The Hund's rule coupling strength is calculated in the experimental
\textit{Pbnm} structure from the orbitally averaged spin splitting of
the diagonal on-site MLWF matrix elements:
${J_{\mathrm{H}}=-\sum_{a,s}\mathrm{sgn}(s)\big(h_{aa}^0\big)^s/4}$,
with ${\mathrm{sgn}(s)=+1/-1}$ for ${s=\,\uparrow/\downarrow}$.

The spin-dependent JT coupling parameter is determined from the
eigenvalue splitting of the on-site MLWF matrix as
${\lambda^{s}={\Delta\varepsilon^{s}}/\left(2\vert{Q^x}\vert\right)}$,
where $\Delta\varepsilon$ is evaluated as
${\Delta\varepsilon=\Big[{\big(h_{11}^{0}-h_{22}^{0}\big)^2+\big(2h_{12}^{0}\big)^2}\Big]^{1/2}}$
in the JT $Q^x$ mode distorted structure.

Similar to the hoppings, the JT coupling parameters $\lambda^s$ are
reduced in the GFO distorted structure by a factor of
${(1-\eta_\lambda)}$, where
${\eta_{\lambda}=1-\big(\Delta\varepsilon^\uparrow_{(\textit{Pbnm})}/\lvert{\bi{Q}}\rvert\big)/\big(\Delta\varepsilon^\uparrow/Q^x\big)}$,
with ${\lvert\bi{Q}\rvert=\sqrt{(Q^x)^2+(Q^z)^2}}$.

\subsection{Interaction parameters}

As it was shown in Ref.~\cite{2012-06//franchini/kovacik//kresse}, the
$U_\mathrm{W}$ interaction parameter can be parameterized either by
mapping the el-el interaction on the difference between the majority
and minority spin on-site matrix elements and suitably introducing an
appropriate correction to the JT splitting
$\Delta \lambda_\mathrm{W}^{(J)}$, or by mapping on the splitting
between the occupied and unoccupied $e_g$ bands with an appropriate
correction to the Hund's coupling $\Delta J_\mathrm{W}^{(\lambda)}$.
Here, we use the latter approach, which is described as follows.

The effective Hubbard parameter $U_\mathrm{W}^{(\lambda)}$ in the MLWF
basis is determined as a correction to the JT induced gap (controlled
by $\lambda^\uparrow$) in the non-interacting (PBE) case.  It is
calculated as
${U_\mathrm{W}^{(\lambda)}=\big(\Delta\varepsilon^\uparrow-\Delta\varepsilon^\uparrow_{(\mathrm{PBE})}\big)/\Delta{n}^\uparrow}$,
where $\Delta\varepsilon^\uparrow$ is the eigenvalue splitting of the
Hamiltonian on-site matrix for a particular beyond-PBE treatment of
the XC functional, $\Delta\varepsilon^\uparrow_{(\mathrm{PBE})}$ its
corresponding value at the PBE level and $\Delta{n}^\uparrow$ the
eigenvalue splitting of the majority occupation matrix in the MLWF
basis (all evaluated in the experimental \textit{Pbnm} structure). The
observation that both the on-site part of the Hamiltonian and the
occupation matrix can be diagonalized by the same unitary
transformation was employed in the formulation.

Since the correlation-induced increase of the spin-splitting is only
partially covered by the one-parameter TB el-el term
$U_\mathrm{W}^{(\lambda)}$, it can be corrected by introducing an
empirical correction to the Hund's rule coupling:
${\Delta
  J_\mathrm{W}^{(\lambda)}=J_\mathrm{H}-J_\mathrm{H(PBE)}-U_\mathrm{W}^{(\lambda)}/4}$.

\section{Methodology: {\em Ab initio} calculations}\label{sec:ab}

Spin polarized DFT calculations were performed using the Vienna
\emph{ab initio} simulation package
(VASP)~\cite{1993-11//kresse/hafner,1996-07//kresse/furthmuller},
without inclusion of spin-orbit coupling.  Three types of XC
functional treatment were employed: (1) the standard GGA with the
parameterization of Perdew-Burke-Ernzerhof (PBE)
~\cite{1996-10//perdew/burke/ernzerhof}; (2) the screened hybrid DFT
following the recipe of Heyd, Scuseria, and Ernzerhof (HSE)
~\cite{2003-05//heyd/scuseria/ernzerhof,2006-06//heyd/scuseria/ernzerhof},
involving the inclusion of 1/4 of the exact Hartree-Fock exchange in
the PBE XC functional; and (3) the GW method~\cite{1965-08//hedin},
where the XC contributions are directly accounted for from the
self-energy. We have adopted a single shot \gowo\ procedure which, at
a relatively moderate computational cost, generally leads to a
significant improvement of the electronic properties with respect to
standard DFT and hybrid functionals. Wavefunctions of the converged
PBE calculation were used as a starting point in the evaluation of the
Green's function G$_0$ and the fixed screened exchange
W$_0$~\cite{2008-09//paier/marsman/kresse,2010-02//franchini/sanna//kresse}.

The one-particle Kohn-Sham orbitals are computed using
projected-augmented-wave (PAW)
pseudopotentials~\cite{1994-12//blochl,1999-01//kresse/joubert}, with
the rare-earth $f$ states frozen in the core (except for La).  The
$3s$ and $3p$ semicore states of Mn, as well as the $5s$ and $5p$
semicore states of $R$, were treated as valence, except for Eu and Gd
where the $5s$ semicore states are excluded from the valence. For
oxygen we have used the soft potential. Integrations in reciprocal
space were carried out over a regular $\Gamma$-centered
${{6}\times{6}\times{4}}$ $k$-point mesh, except for \gowo\ where a
reduced setting of ${{4}\times{4}\times{4}}$ was adopted.  The
plane-wave energy cutoff was set to 400 eV for PBE and \gowo. After
testing the influence of the energy cutoff on the HSE tight binding
parameters, a value of 300 eV was used in all HSE calculations to
reduce the computational cost. The total number of bands was increased
to 320 in the \gowo\ runs.  This value leads to sufficiently well
converged band gap (within 0.1-0.2 eV) but a larger value would be
needed to better describe the highest unoccupied $e_{g}$ manifold.
Unfortunately, the inclusion of a larger number of bands would result
in a prohibitive computational cost.

Ground state electronic, optical and magnetic properties were
calculated for the experimental \textit{Pbnm} structure in the A-AFM
order.  By mapping the total energy differences among different
magnetic configurations to the Heisenberg Hamiltonian, exchange
coupling parameters were evaluated within the HSE approach.  With the
so determined exchange parameters, an estimate for the N\'eel
temperature was computed via Monte Carlo simulations (MC), employing
the Metropolis algorithm~\cite{1953-06//metropolis/rosenbluth//teller}
and using the Mersenne twister for the random number
generation~\cite{1998-01//matsumoto/nishimura}.  Finally, the TB
parameters were extracted from the Hamiltonian matrix elements in the
basis of the MLWFs, constructed from the \emph{ab initio}
wavefunctions in the A-AFM experimental and $Q^x$ mode JT structures
with the VASP2WANNIER90
interface~\cite{2012-06//franchini/kovacik//kresse}.  For practical
reasons, the $f$ states of La were pushed away from the Mn $e_g$
energy window in the PBE calculations by applying $U=10$~eV, following
the recipe of Dudarev \emph{et
  al.}~\cite{1998-01//dudarev/botton//sutton}.

\section{Results and discussion}\label{sec:res}

In this section we present the outcomes of the combined \emph{ab
  initio} and model Hamiltonian analysis. First we discuss the ground
state electronic structure, MLWFs and dielectric properties as derived
from the \emph{ab initio} PBE, HSE and \gowo\ calculations. Then we
will focus on the detailed explanation of the origin of the A-type AFM
ordering by mapping the HSE total energies onto a Heisenberg
Hamiltonian and computing the ordering temperature $T_\text{N}$ from
Monte Carlo simulations. Finally, an extended section will be
dedicated to the TB results.

\subsection{Electronic and dielectric properties}\label{ssec:res-el}

\begin{figure*}[t!]
  \centering
  \includegraphics{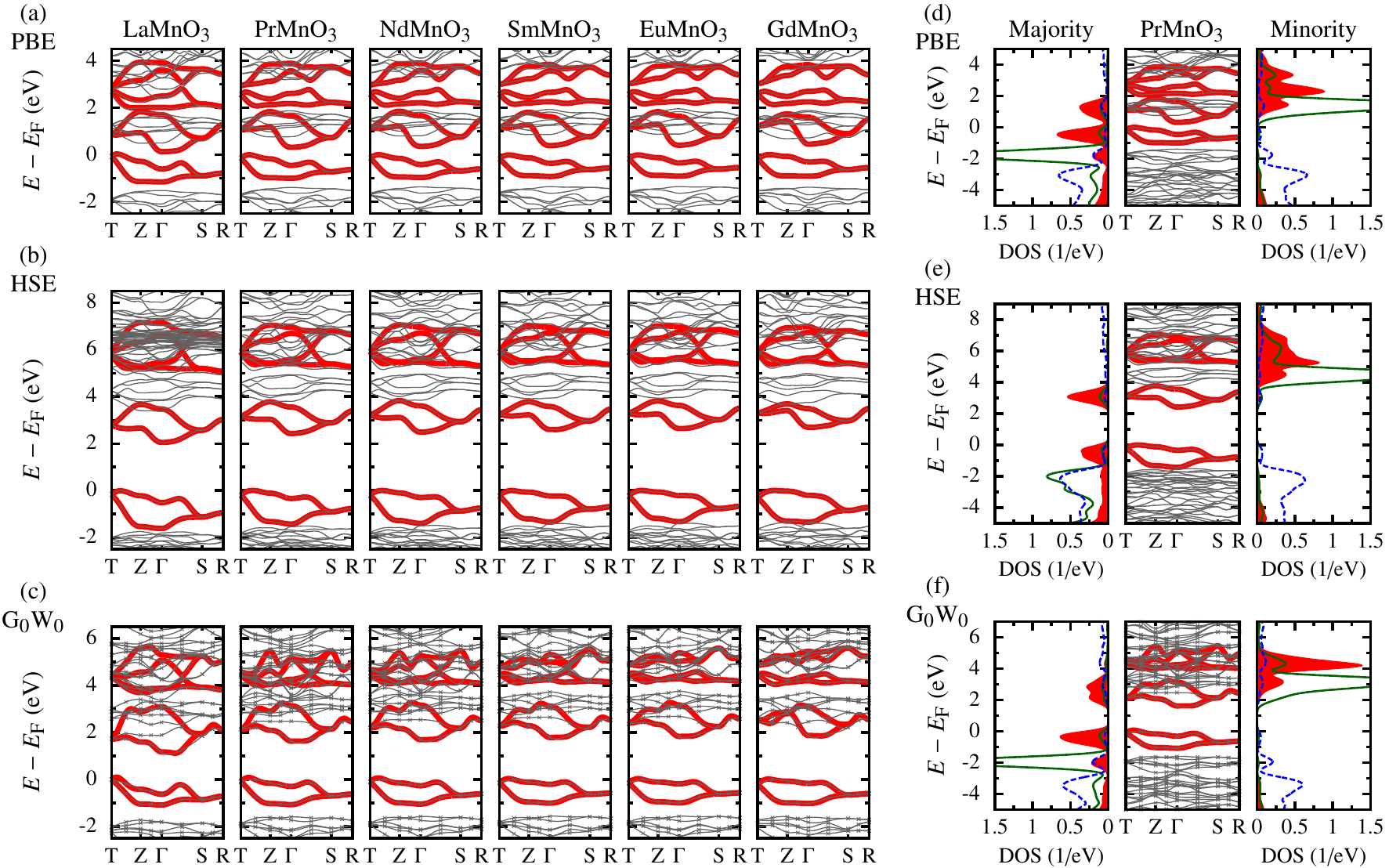}
  \caption{(color online) The \emph{ab initio} (thin lines) and $e_g$
    character MLWF (thick lines) band structure of $R$MnO$_{3}$ ($R=$
    La, Pr, Nd, Sm, Eu, Gd); for (a) PBE (b) HSE and (c) \gowo. Band
    structure and associated normalized projected density of states
    (PDOS) for PrMnO$_{3}$: Mn $e_{g}$ (filled areas under curve), Mn
    $t_{2g}$ (solid lines) and O $p$ (dashed line) for (d) PBE (e) HSE
    and (f) \gowo. Left/right PDOS correspond to local
    majority/minority Mn sites while O $p$ PDOS is calculated as an
    average over all O sites.}
  \label{fig:rmno3-all}
\end{figure*}

Figure~\ref{fig:rmno3-all}~(a), (b) and (c) depict the calculated band
structures at the PBE, HSE and \gowo\ level, respectively, for
$R$MnO$_3$ ($R=$ La, Pr, Nd, Sm, Eu and Gd) along with the
corresponding characteristic MLWFs bands of predominantly $e_{g}$
character.
It is seen that upon $R$ substitution, the features of the $e_g$
character bands do not exhibit substantial differences. Consequently,
the electronic properties, including screening effects, could be
expected to remain almost unchanged over the series.  The band
structures depict an insulating state with an indirect energy gap.  As
a representative example of all compounds in the series, we show in
Figs.~\ref{fig:rmno3-all}(d)-(f) the band structure and associated
projected density of states (PDOS) of PrMnO$_3$.  The PDOS is shown
for the Mn $e_{g}$/$t_{2g}$ and O $p$ character.  The overall bonding
picture resembles closely the one of
LaMnO$_3$~\cite{2012-06//franchini/kovacik//kresse,2012-12//he/franchini,2012-05//he/chen//franchini}:
the indirect band gap is opened between the lower laying $e_g$ states,
there is a strong hybridization between Mn $d$ and O $p$ states, and
an appreciable intermixing between Mn $e_{g}$ and $t_{2g}$ states is
observed, in particular around the band gap.

The band gaps of $R$MnO$_{3}$ and the local magnetic moments at the
Mn$^{3+}$ sites with the different levels of exchange-correlation
treatment are presented in Tab.~\ref{tab:mm-bg}.
\begin{table}[!t]
  \centering
  \caption{The values of indirect (smallest direct) band gap and magnetic
    moment at the Mn$^{3+}$ sites of $R$MnO$_{3}$ series.} 
      \begin{tabularx}{246pt}{c@{\enskip}cYYcY}
        \hline\hline
	&\multicolumn{3}{c}{Band gap (eV)}\\
        \cline{2-5}
	&\multicolumn{1}{c}{PBE} & \multicolumn{1}{c}{HSE} & \multicolumn{1}{c}{\gowo} & \multicolumn{1}{c}{Expt.}\\
	\hline
LaMnO$_3$ & 0.13 (0.56) & 2.06 (2.48) & 1.15 (1.49) &  1.7$^\text{a}$, 1.9$^\text{b}$, 2.0$^\text{c}$\\
PrMnO$_3$ & 0.32 (0.72) & 2.43 (2.74) & 1.63 (1.81) &  1.75$^\text{d}$, 2.0$^\text{e}$ \\
NdMnO$_3$ & 0.36 (0.74) & 2.49 (2.78) & 1.70 (1.86) &  1.75$^\text{d}$, 1.78$^\text{f}$\\
SmMnO$_3$ & 0.40 (0.75) & 2.61 (2.80) & 1.79 (1.92) &  1.82$^\text{f}$\\
EuMnO$_3$ & 0.42 (0.75) & 2.65 (2.85) & 1.84 (1.96) &  \\
GdMnO$_3$ & 0.45 (0.75) & 2.70 (2.89) & 1.87 (1.99) &  2.0$^\text{g}$, 2.9$^\text{h}$\\
        \hline
        \hline
        &\multicolumn{3}{c}{Magnetic moment ($\mu_{B}$)}\\
        \cline{2-5}
        &\multicolumn{1}{c}{PBE} & \multicolumn{1}{c}{HSE} & \multicolumn{1}{c}{\gowo} & \multicolumn{1}{c}{Expt.}\\
        \hline
LaMnO$_3$ &  3.49 & 3.72 & 3.40 & 3.4$^\text{i}$, 3.87$^\text{j}$, 3.65$^\text{k}$\\
PrMnO$_3$ &  3.49 & 3.71 & 3.40 & 3.5$^\text{l}$\\
NdMnO$_3$ &  3.49 & 3.71 & 3.40 & 3.22$^\text{m}$\\
SmMnO$_3$ &  3.49 & 3.71 & 3.40 & 3.3$^\text{n}$\\
EuMnO$_3$ &  3.49 & 3.71 & 3.40 & \\
GdMnO$_3$ &  3.47 & 3.70 & 3.38 & \\
        \hline\hline
      \end{tabularx}

% References
\medskip
$^\text{a\,}$Ref.~\cite{1995-05//saitoh/bocquet//takano};
$^\text{b\,}$Ref.~\cite{1997-06//jung/kim//chung};
$^\text{c\,}$Ref.~\cite{2004-03//kruger/schulz//rubhausen};
$^\text{d\,}$Ref.~\cite{2006-06//kim/moon//noh};
$^\text{e\,}$Ref.~\cite{2006-09//sopracase/gruener//soret};
$^\text{f\,}$Ref.~\cite{2010-12//shetkar/salker};
$^\text{g\,}$Ref.~\cite{2010-05//wang/li//zhang};
$^\text{h\,}$Ref.~\cite{2013-05//negi/dixit//srivastava};
$^\text{i}\,$Ref.~\cite{1996-06//hauback/fjellvag/sakai};
$^\text{j}\,$Ref.~\cite{1996-12//moussa/hennion//revcolevschi};
$^\text{k}\,$Ref.~\cite{1997-06//huang/santoro//greene};
$^\text{l}\,$Ref.~\cite{2006-06//laverdiere/jandl//iliev};
$^\text{m}\,$Ref.~\cite{2000-02//munoz/alonso//fernandez-diaz};
$^\text{n}\,$Ref.~\cite{2011-05//oflynn/tomy//balakrishnan}.
  \label{tab:mm-bg}
\end{table}
The magnetic moment at the Mn$^{3+}$ sites remains basically unaltered
along the $R$ series, while a general trend of the band gap increase
from La to Gd is seen at all XC levels.  The overall increment in the
direct band gap is of about 0.4 and 0.5~eV at the HSE and \gowo\
level, respectively, and in the indirect band gap of about 0.7~eV in
both cases. The experimental data, not available for EuMnO$_3$, do not
show a clear trend but are generally in line with the \gowo\
expectations for the direct band gap. Although still capturing the
insulating nature, PBE results in much too small values of the band
gap as expected. On the other hand, the HSE values appear too
overestimated. This is likely due to the amount of exact exchange
incorporated in the HSE functional. In the present study, we have used
the standard 0.25
compromise~\cite{2003-05//heyd/scuseria/ernzerhof,2006-06//heyd/scuseria/ernzerhof}. However,
recent systematic studies on the role of the mixing parameter on the
physical properties of perovskites have indicated that a lower
fraction should be used (0.1-0.15), to achieve a more consistent
picture~\cite{2012-12//he/franchini,2014-06//franchini}.

\begin{figure}[!t]
  \centering
  \includegraphics{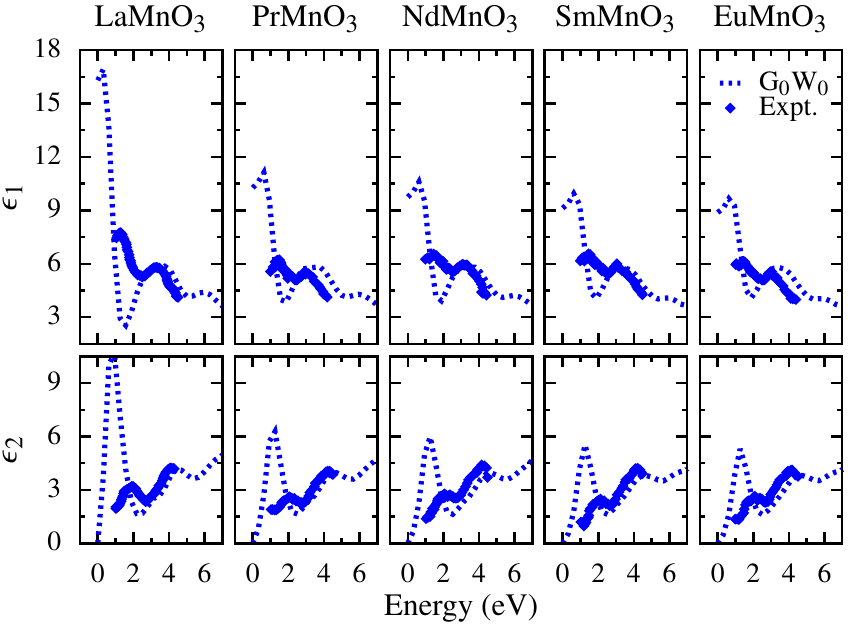}
  \caption{(color online) The real ($\epsilon_{1}$) and imaginary
    ($\epsilon_{2}$) part of the dielectric function. The dotted lines
    indicate the \gowo\ data and the filled diamonds correspond to the
    experimental data taken from
    Ref.~\cite{2010-07//moskvin/makhnev//balbashov}.}
  \label{fig:dielec}
\end{figure}

The dielectric function measured in an energy range between 0.5 to 5.5
eV shows two intensive, broad optical features peaked at approximately
2~eV and 4.5~eV for LaMnO$_{3}$ (Fig.~\ref{fig:dielec}).
For the other $R$MnO$_{3}$ compounds, the intensive broad peak is
positioned at \mbox{$\approx{2.2}$} eV. While the authors in
Refs.~\cite{2000-05//ahn/millis,2004-10//kovaleva/boris//keimer}
assign the peaks to ${d\hbox{-}d}$ charge transfer excitations, the
authors of Ref.~\cite{2010-07//moskvin/makhnev//balbashov} argue that
the peaks are due to the interplay of both ${p\hbox{-}d}$ and
${d\hbox{-}d}$ transitions.  These experimental results are in line
with the measurements of
Kim~\emph{et~al.}~\cite{2009-08//kim/min}. The \gowo\ results capture
well the double peak structure, but the intensity of the first peak
and the zero frequency value of the real part of the dielectric
function $\epsilon_{1}$, which identifies the macroscopic dielectric
constant $\epsilon_\infty$, is about two times larger than the
experimental one. A better agreement with experiment could possibly be
achieved by increasing the number of bands, the $k$-points sampling
and by treating the screened exchange at beyond-PBE level (i.e.,
within a fully self-consistent GW framework) but this is beyond the
scope of the present study (the corresponding calculation would be
computationally very demanding)~\cite{2014-01//he/franchini} and will
be addressed in a future article~\cite{ergonenc}.

\subsection{Magnetic properties}\label{ssec:res-mag}

\begin{figure}[b!]
  \includegraphics{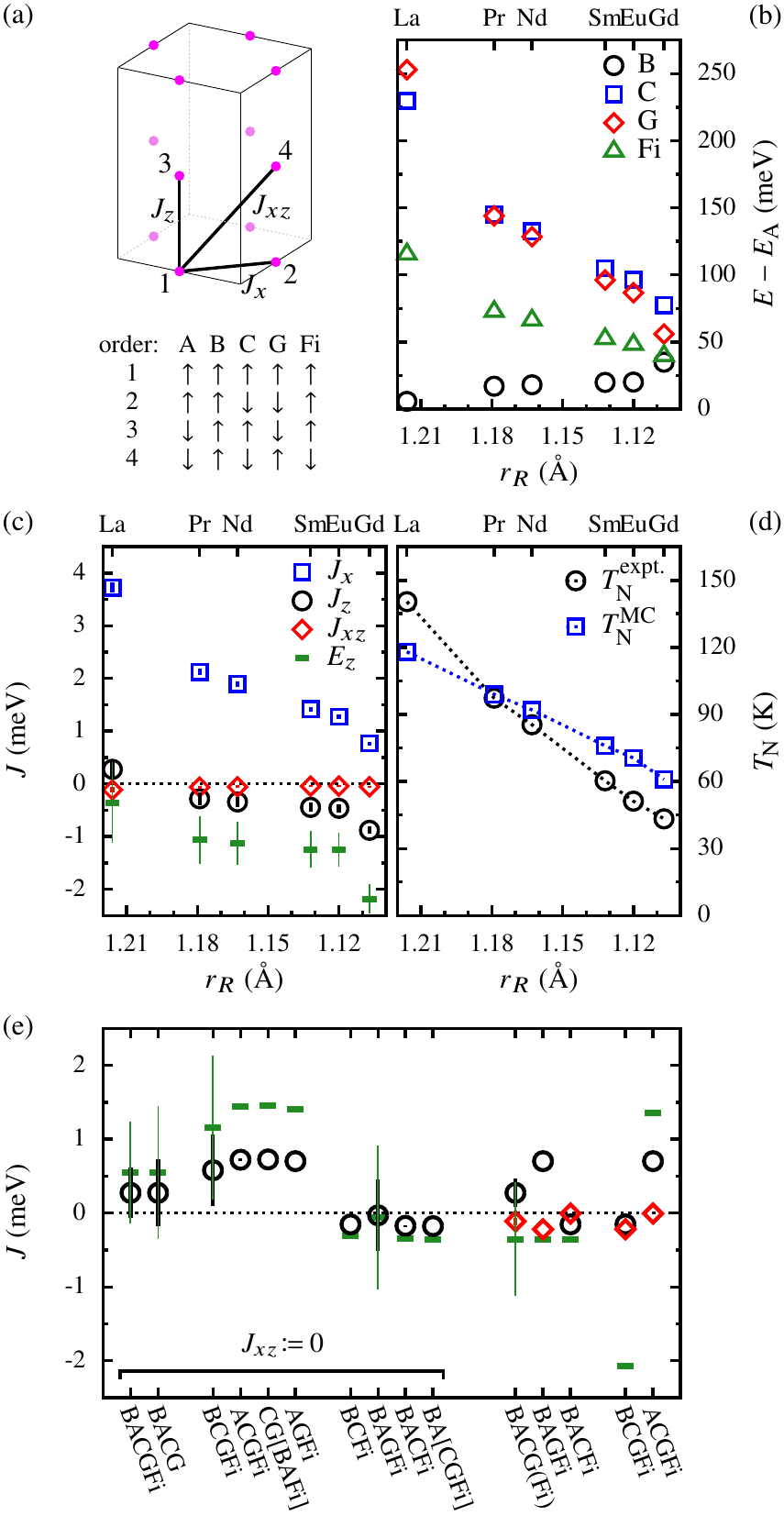}
  \caption{(color online) (a) Schematic representation of the exchange
    interactions $J_x$, $J_z$ and $J_{xz}$ in the unit cell of
    $R$MnO$_3$, together with the spin orientations in the A, B, C, G
    and Fi magnetic configurations. (b) The total energy of the
    different magnetic configurations relative to the A ordering. (c)
    Exchange interactions calculated by linear least-square fit of
    Eqs.~\eqref{eq:EA} to \eqref{eq:EFi} and the out-of-plane
    interlayer interaction energy $E_{z}=2J_{z}+8J_{xz}$.  Errors on
    the estimation of the corresponding parameter are shown as
    vertical bars.  (d) The N\'eel temperature, experimental
    $T_\text{N}^{\text{expt.}}$~\cite{2003-08//kimura/ishihara//tokura}
    and calculated $T_\text{N}^{\text{MC}}$ via MC simulations using
    the exchange interactions shown in (c).  (e) $J_z$, $J_{xz}$ and
    $E_z$ in LaMnO$_3$ for different sets of magnetic
    configurations. For sets yielding identical values of $J_z$, we
    adopt a notation of choose one from the list in square brackets.
    The BACG(Fi) data depicts both BACGFi (with error bars) and BACG
    (without error bars). Legend is identical to (c).}
  \label{fig:mag}
\end{figure}

We further analyze the magnetic properties of the $R$MnO$_3$ compounds
in terms of the exchange interactions $J_{ij}$ between sites $i$ and
$j$, obtained by mapping the total energy of different magnetic
configurations on the Heisenberg Hamiltonian
\begin{equation}\label{eq:hh}
  H=-\frac{1}{2}\sum_{i
    \neq j} J_{ij}\,\bi{S}_{i}\cdot\bi{S}_{j},
\end{equation}
for ${\lvert{\bi{S}_i}\rvert=\lvert{\bi{S}_j}\rvert=2}$, with positive
and negative values of $J_{ij}$ corresponding to FM and AFM coupling,
respectively.  In the four formula unit cell there are three exchange
interactions that can be extracted: the in-plane nearest neighbor
$J_{x}$, the out-of-plane nearest neighbor $J_{z}$ and second nearest
neighbor $J_{xz}$; where the subscripts are a shorthand notation of
the direction connecting the sites in the pseudo-cubic axes frame [see
Fig.~\ref{fig:mag}(a)]. Determining interactions between further
neighbors would require a larger supercell.  While often only the
first two parameters are taken into
consideration~\cite{2004-02//munoz/harrison/illas,2005-12//evarestov/kotomin//maier},
it was reported that the A-AFM order in LaMnO$_3$ can be seen as a
competition between a weakly FM $J_{z}$ and a weakly AFM $J_{xz}$
coupling~\cite{1996-06//solovyev/hamada/terakura}.  As it was
discussed previously, simple treatments of the exchange-correlation
functional (such as PBE) were shown to be inadequate in providing a
good prediction of the magnetic properties/interactions for
perovskites and in general for transition metal
oxides~\cite{2008-07//yamauchi/freimuth//picozzi,
  2011//archer/chaitanya//mujumdar}. Although the exchange
interactions in LaMnO$_3$ calculated using hybrid functionals were
found to be largely dependent on the choice of the particular hybrid
functional, the A-AFM order is consistently predicted to be the
magnetic ground state~\cite{2004-02//munoz/harrison/illas}. We
therefore base our analysis on the total energies calculated using the
HSE functional, that has been already employed successfully in
combination with the Monte Carlo method to predict the magnetic
ordering temperature in transition metal
perovskites~\cite{2001//franchini/archer//sanvito}.

We also note that a recent study has shown that the magnetic
properties of the later members of the manganite series \mbox{($R=$ Tb
  to Lu)} are not well described by a standard Heisenberg model with
pairwise bilinear interactions, and that additional biquadratic or
four-spin ring exchange interactions need to be
considered~\cite{2015-04//fedorova/ederer//scaramucci}. However, for
the larger rare earth cations considered in this study, the Heisenberg
model is expected to provide a sufficiently accurate description.
  
The total energy was calculated for the five symmetry inequivalent
magnetic configurations compatible with the unit cell. These include:
the ferromagnetic order (B); the three distinct antiferromagnetic
configurations A-AFM (A), C-AFM (C) and G-AFM (G); and the single
non-degenerate ferrimagnetic state (Fi), as depicted in
Fig.~\ref{fig:mag}(a).
For brevity, the shorthand notation in
parentheses~\cite{1955-10//wollan/koehler} will be used in this
section to denote the total energies of the corresponding magnetic
configuration. These, by using Eq.~\eqref{eq:hh}, are expressed as:
\begin{subequations}
\begin{eqnarray}
  E_\text{A} &=& E_0-32J_{x}+16J_z+64J_{xz},\label{eq:EA}\\
  E_\text{B} &=& E_0-32J_{x}-16J_z-64J_{xz},\label{eq:EB}\\
  E_\text{C} &=& E_0+32J_{x}-16J_z+64J_{xz},\label{eq:EC}\\
  E_\text{G} &=& E_0+32J_{x}+16J_z-64J_{xz},\label{eq:EG}\\
  E_\text{Fi} &=& E_0. \label{eq:EFi}
\end{eqnarray}
\end{subequations}
where $E_0$ is a fitting constant in unit of energy that should
correspond to the energy of the paramagnetic state. The total energies
relative to $E_\text{A}$ are plotted in Fig.~\ref{fig:mag}(b). For all
members of the $R$ series, the A ordering yields the lowest energy
among the five considered magnetic configurations.  While the
difference in the total energy from the C or G orders on one side to
the A or B orders on the other side are relatively large in case of
LaMnO$_3$, these differences decrease consistently towards GdMnO$_3$,
following the trend of decreasing $T_\text{N}$. We note that
$(E_\text{C}-E_\text{A})$ does not differ from
$2(E_\text{Fi}-E_\text{A})$ by more than 2 meV.

The solution to the overdetermined system of equations composed by
Eqs.~\eqref{eq:EA} to \eqref{eq:EFi} is obtained by linear
least-squares fit and the resulting exchange interaction parameters
$J_{x}$, $J_{z}$ and $J_{xz}$ are shown in Fig.~\ref{fig:mag}(c). The
in-plane $J_{x}$ interaction is FM throughout the whole $R$ series,
monotonously and strongly decreasing from La to Gd. The out-of-plane
$J_{z}$ interaction exhibits a similar trend, however, the weakly FM
$J_{z}$ for $R=$ La becomes weakly AFM from $R=$ Pr on. The stability
of A order is finally determined by the out-of-plane interaction
energy $E_{z}=2J_{z}+8J_{xz}$ being negative for all members of $R$
series, including La. There, however, the error on the estimation of
$E_{z}$ is large enough to reach the FM region.  By employing the
Monte Carlo method, the exchange interaction parameters are used to
calculate the N\'eel temperature presented in Fig.~\ref{fig:mag}(d),
generally leading to very good agreement with the experimental
values~\cite{2003-08//kimura/ishihara//tokura}.

\begin{figure*}[t!]
\includegraphics{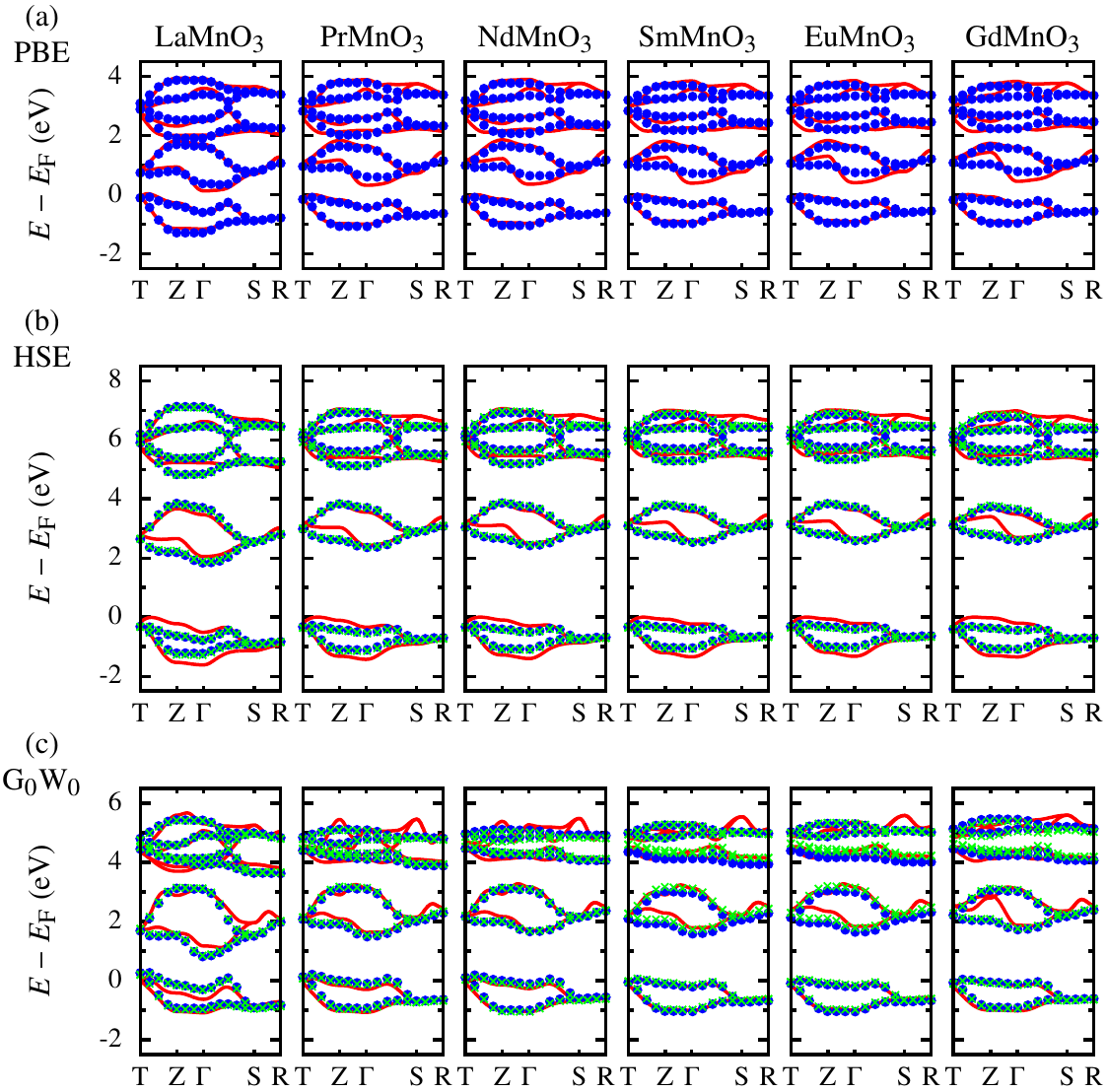}
\caption{(color online) The TB bands calculated according to Model~1
  (filled circles) and Model~2 (crosses) with the MLWFs bands (solid
  lines) in the background.}
\label{fig:tb}
\end{figure*}

We note that for the case of LaMnO$_3$ it is of particular importance
to include the out-of-plane second neighbor interaction in the
model. Furthermore, the calculated exchange interactions are
particularly sensitive to the choice of the subset $M$ of magnetic
configurations to include in the system of equations [see
Fig.~\ref{fig:mag}(e)]. By setting $J_{xz}=0$ and operating
Eqs.~\eqref{eq:EA} to \eqref{eq:EFi}, it is easy to show that $J_z$
can be calculated from any of the following expressions:
\begin{subequations}
\begin{empheq}
 [ left ={J_z = \displaystyle\frac{1}{32} \empheqlbrace\enskip}]{align}
  &  (E_\text{A}+E_\text{G}-2E_\text{Fi}) &:\quad& \{\text{AGFi}\} \in M\label{eq:jz1}\\
  &  (2E_\text{Fi}-E_\text{B}-E_\text{C}) &:\quad& \{\text{BCFi}\} \in M\label{eq:jz2}\\
  &  (E_\text{A}-E_\text{B}) &:\quad& \{\text{AB}\} \in M\label{eq:jz3}\\
  &  (E_\text{G}-E_\text{C}) &:\quad& \{\text{CG}\} \in M\label{eq:jz4}.
\end{empheq}
\end{subequations}
For $R =$ Pr to Gd, the $J_z$ calculated from Eqs.~\eqref{eq:jz1}
to~\eqref{eq:jz4} is always negative, whereas in LaMnO$_3$ its sign
becomes positive when using Eqs.~\eqref{eq:jz1}
and~\eqref{eq:jz4}. Using all five energy points (BACGFi) the linear
least-squares fit to Eqs.~\eqref{eq:EA} to \eqref{eq:EFi} yields a FM
$J_z$ whose magnitude is the average obtained from Eqs.~\eqref{eq:jz1}
and~\eqref{eq:jz2}, but with an error in the estimation of the
parameter large enough to turn it AFM. An out-of-plane FM coupling is
obtained when either AGFi or CG are present in $M$, which is
inconsistent with the experimentally observed magnetic ground state
(sets BCGFi, ACGFi, BCG, ACG, CGFi and AGFi). Conversely, when either
BCFi or AB are present in $M$ (sets BCFi, BAGFi, BACFi, BAC, BAG, and
BAFi), the system of equations yields a negative $J_z$ in agreement
with experiments. However, that would be the equivalent of removing
inconvenient data to adjust it to a desired outcome, when actually
these inconsistencies can be ascribed to the fact that without
$J_{xz}$ the expansion of the Heisenberg Hamiltonian is
incomplete. Including $J_{xz}$ as a third interaction parameter, only
Eqs.~\eqref{eq:jz1} and~\eqref{eq:jz2} hold. The out-of-plane magnetic
coupling is driven by the previously defined interaction energy $E_z$,
which can be regarded as the effective out-of-plane exchange
interaction. As shown in Fig.~\ref{fig:mag}(e), provided that both A
and B are present in $M$, $E_z$ is not only negative but rather
insensitive to the configuration of the subset, in spite of the
pronounced differences obtained in the particular values of $J_z$ and
$J_{xz}$. This is so because $E_z$ is calculated as
\begin{subequations}
\begin{empheq}
 [ left ={\!\!E_z = \displaystyle\frac{1}{16} \empheqlbrace\enskip} ]{align}
  &  (E_\text{A}-E_\text{B}) &\!\!\!\!\!\!:\ \ & \{\text{AB}\} \in M\label{eq:ez1}\\
  &  (4E_\text{Fi}-2E_\text{B}-E_\text{C}-E_\text{G}) &\!\!\!\!\!\!:\ \ &
  \text{A} \notin M\label{eq:ez2}\\
  &  (2E_\text{A}+E_\text{C}+E_\text{G}-4E_\text{Fi}) &\!\!\!\!\!\!:\ \ &
  \text{B} \notin M\label{eq:ez3}.
\end{empheq}
\end{subequations}
When either A or B are not members of $M$, Eqs.~\eqref{eq:ez2}
and~\eqref{eq:ez3} would be equivalent to Eq.~\eqref{eq:ez1} if the
following identity, stemming  from Eqs.~\eqref{eq:EA} to \eqref{eq:EFi}, is verified:
\mbox{$E_\text{A}+E_\text{B}+E_\text{C}+E_\text{G}=4E_\text{Fi}$}. However,
not only the Heisenberg model is itself an approximation but
also the \emph{ab initio} total energies are not exempt from errors
due to various approximations affecting the calculations. Since the
out-of-plane exchange parameters are of comparable magnitude to these
errors (in the order of meV), minor deviations in the previous
equality lead to the observed large differences in $J_z$, $J_{xz}$ and
$E_z$. This is the reason why it is advisable and often necessary
to extract the exchange parameters from as large sets of magnetic
configurations as possible.

\subsection{Tight binding}\label{ssec:res-tb}

\begin{figure}[b!]
\includegraphics{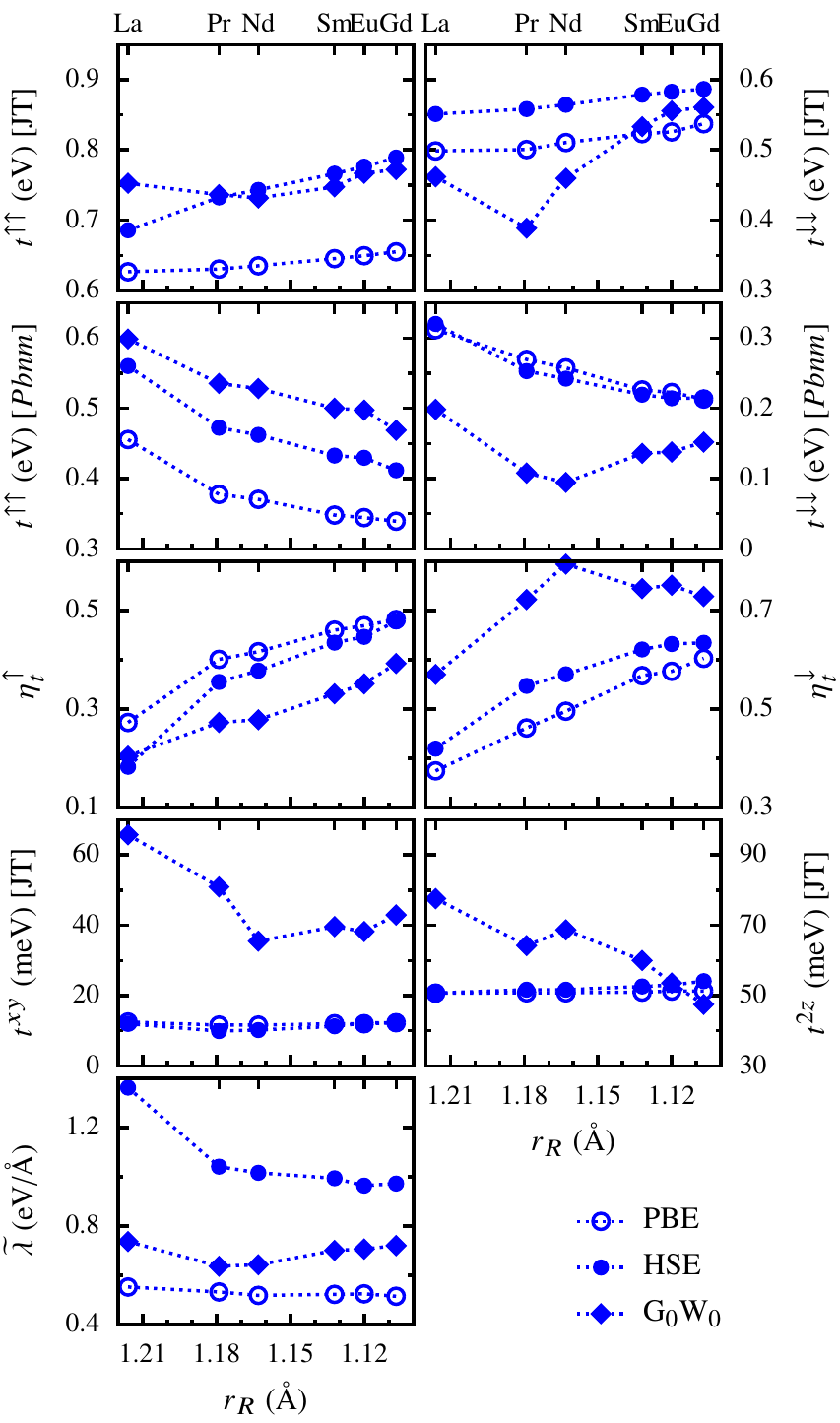}
\caption{Dependence of the TB hopping related model parameters on
  $r_{R}$. The nearest neighbor hopping parameters
  $t^{\uparrow\uparrow}$ and $t^{\downarrow\downarrow}$ are shown in
  the JT($Q^x$) and experimental \textit{Pbnm} crystal structure,
  which are used to determine the GFO-induced reduction factors
  $\eta^\uparrow$ and $\eta^\downarrow$. As next, the further neighbor
  hopping parameters $t^{xy}$ and $t^{2z}$ and the JT-induced
  splitting in the non-diagonal elements of the nearest neighbor
  in-plane hopping matrix $\widetilde{\lambda}$ are shown.}
\label{fig:modelparhop}
\end{figure}

\begin{figure}[b!]
\includegraphics{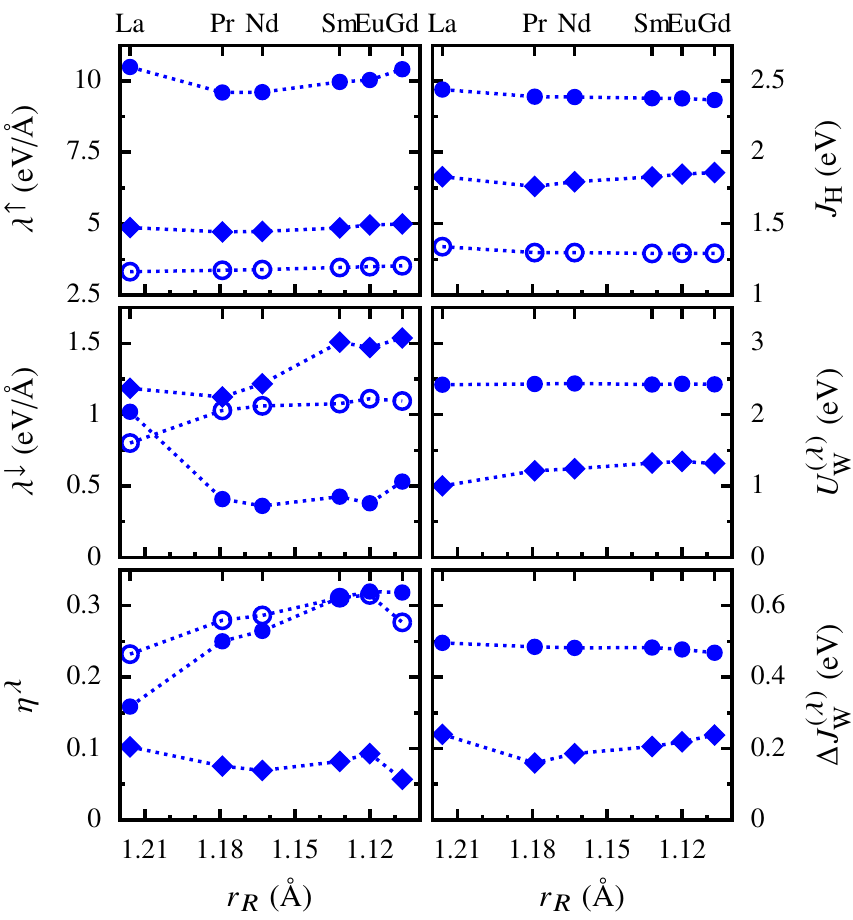}
\caption{Dependence of the TB on-site model parameters on $r_{R}$: JT
  coupling parameters $\lambda^\uparrow$, $\lambda^\downarrow$ and the
  GFO-induced reduction parameter of the JT coupling $\eta^\lambda$,
  Hund's coupling $J_\mathrm{H}$, the mean-field Hubbard parameter
  $U_\mathrm{W}^{(\lambda)}$ and the correction to the Hund's coupling
  $\Delta{J}_W^{(\lambda)}$. For legend see
  Fig.~\ref{fig:modelparhop}.}
\label{fig:modelparonsite}
\end{figure}

We remind that we performed two types of TB parameterization: Model~1
and Model~2. In Model~1, the term $\hat{H}_\mathrm{el-el}$ is not
considered, the el-el interaction is implicitly accounted for in the
HSE and \gowo\ hopping, JT- and GFO-induced parameters, which will
differ from the corresponding PBE values. In Model~2, the
modifications due to the beyond-PBE methods are treated as a
perturbation to the ``noninteracting'' PBE description by explicitly
considering the $\hat{H}_\mathrm{el-el}$ term in the mean-field
approximation. The band structures obtained with these sets of TB
parameters compared with the corresponding MLWFs bands are shown in
Fig.~\ref{fig:tb}.  The individual TB parameters are shown in
Figs.~\ref{fig:modelparhop} and~\ref{fig:modelparonsite} and are
presented in detail in Tab.~\ref{tab:tb}.

In general, very good qualitative agreement can be seen between the
features of the TB and MLWF bands (Fig.~\ref{fig:tb}). Moreover,
almost no difference is found between the bands calculated with the TB
parameters using Model~1 and Model~2. While the match for LaMnO$_3$ at
PBE level (for which the procedure was originally developed in
Ref.~\cite{2010-06//kovacik/ederer}) is very good, deviations for the
lowest unoccupied $e_g$ character band increase along the $R$
series. This is not surprising considering that the progressively
stronger GFO distortion makes the assumption of the individual
structural distortions acting independently less valid. Nevertheless,
the root mean square and maximum deviation between the band and
$k$-point averaged sets of eigenvalues for the TB and MLWF bands are
typically around very acceptable values: 0.15 and 0.5 eV,
respectively. The quantitative deviations observed in the \gowo\ local
minority bands can be, on the other hand, attributed to difficulties
in achieving well-converged results at the \gowo\ level.

In the following we analyze in detail how the hopping and on-site TB
parameters are affected along the $R$ series at different levels of
the XC functional treatment.

\begin{table*}
  \centering
  \caption{The tight binding parameters. The hopping parameters
    $t^{\uparrow\uparrow}$ and $t^{\downarrow\downarrow}$
    for the JT structure in Fig.~\ref{fig:modelparhop} correspond to the
    $t^{\uparrow\uparrow}$ and $t^{\downarrow\downarrow}$ presented in
    this table.}
  \begin{tabularx}{510pt}{c|l@{\quad}lYYYYYYY@{\quad}lYYYY@{\quad}lYYYY}
    \hline\hline
    \multicolumn{3}{c}{}&\multicolumn{7}{c}{Hopping
      parameters}&&\multicolumn{4}{c}{On-site
      parameters}&&\multicolumn{4}{c}{Model~2} \\
    \cline{4-15}\cline{17-20}\multicolumn{20}{c}{}\\[-10pt]
    \multicolumn{3}{c}{}&$t^{\uparrow\uparrow}$ &
    $t^{\downarrow\downarrow}$ & $\widetilde{\lambda}$ & $t^{xy}$ & $t^{2z}$ & $\eta_t^{\uparrow}$ & $\eta_t^{\downarrow}$&&
    $J_\mathrm{H}$ & $\lambda^{\uparrow}$ & $\lambda^{\downarrow}$ &
    $\eta_{\lambda}$&&
    $\Delta \varepsilon^{\uparrow}$  & $\Delta{n}^{\uparrow}$ &
    $U_\mathrm{W}^{(\lambda)}$ & $\Delta{J}_\mathrm{W}^{(\lambda)}$ \\
    \multicolumn{3}{c}{}&(meV)&(meV)&(eV/\AA)&(meV)&(meV)&&&&
    (eV)&(eV/\AA)&(eV/\AA)&&&(eV)&&(eV)&(eV)\\
    \hline
    \multirow{6}{*}{\rotatebox{90}{PBE}}
    &LaMnO$_3$&
    & 627 & 499 & 0.55 & 12 & 51 & 0.27 & 0.37 &
    &  1.34 & 3.31 &  0.80 & 0.23 &
    &  0.85 & & &
    \\ 
    &PrMnO$_3$&
    & 631 & 500 & 0.53 & 12 & 51 & 0.40 & 0.46 &
    &  1.30 & 3.36 &  1.03 & 0.28 &
    &  1.12 & & &
    \\ 
    &NdMnO$_3$&
    & 635 & 511 & 0.52 & 12 & 51 & 0.42 & 0.50 &
    &  1.30 & 3.39 &  1.06 & 0.29 &
    &  1.15 & & &
    \\ 
    &SmMnO$_3$&
    & 645 & 523 & 0.52 & 12 & 51 & 0.46 & 0.57 &
    &  1.29 & 3.46 &  1.08 & 0.31 &
    &  1.18 & & &
    \\ 
    &EuMnO$_3$&
    & 649 & 526 & 0.52 & 12 & 51 & 0.47 & 0.58 &
    &  1.29 & 3.49 &  1.11 & 0.31 &
    &  1.21 & & &
    \\ 
    &GdMnO$_3$&
    & 655 & 537 & 0.51 & 12 & 51 & 0.48 & 0.60 &
    &  1.29 & 3.52 &  1.10 & 0.28 &
    &  1.26 & & &
    \\ 
    \hline
    \multirow{6}{*}{\rotatebox{90}{HSE}}
    &LaMnO$_3$&
    & 686 & 551 & 1.36 & 12 & 51 & 0.18 & 0.42 &
    &  2.44 & 10.49 &  1.02 & 0.16 &
    &  2.96 &  0.87 &  2.42 &  0.50
    \\ 
    &PrMnO$_3$&
    & 732 & 558 & 1.04 & 10 & 52 & 0.35 & 0.55 &
    &  2.39 & 9.59 &  0.41 & 0.25 &
    &  3.31 &  0.90 &  2.43 &  0.48
    \\ 
    &NdMnO$_3$&
    & 743 & 564 & 1.02 & 10 & 52 & 0.38 & 0.57 &
    &  2.39 & 9.61 &  0.36 & 0.26 &
    &  3.36 &  0.91 &  2.44 &  0.48
    \\ 
    &SmMnO$_3$&
    & 766 & 579 & 0.99 & 11 & 53 & 0.44 & 0.62 &
    &  2.38 & 9.97 &  0.43 & 0.31 &
    &  3.41 &  0.92 &  2.42 &  0.48
    \\ 
    &EuMnO$_3$&
    & 776 & 583 & 0.96 & 12 & 53 & 0.45 & 0.63 &
    &  2.38 & 10.03 &  0.38 & 0.32 &
    &  3.45 &  0.92 &  2.43 &  0.48
    \\ 
    &GdMnO$_3$&
    & 789 & 587 & 0.97 & 12 & 54 & 0.48 & 0.63 &
    &  2.37 & 10.41 &  0.53 & 0.32 &
    &  3.50 &  0.93 &  2.43 &  0.47
    \\ 
    \hline
    \multirow{6}{*}{\rotatebox{90}{\gowo}}
    &LaMnO$_3$&
    & 753 & 462 & 0.74 & 66 & 78 & 0.20 & 0.57 &
    &  1.83 & 4.86 &  1.18 & 0.10 &
    &  1.46 &  0.61 &  1.00 &  0.24
    \\ 
    &PrMnO$_3$&
    & 736 & 389 & 0.64 & 51 & 64 & 0.27 & 0.72 &
    &  1.76 & 4.71 &  1.12 & 0.08 &
    &  2.00 &  0.73 &  1.21 &  0.16
    \\ 
    &NdMnO$_3$&
    & 732 & 460 & 0.64 & 36 & 69 & 0.28 & 0.79 &
    &  1.79 & 4.73 &  1.22 & 0.07 &
    &  2.10 &  0.76 &  1.24 &  0.19
    \\ 
    &SmMnO$_3$&
    & 747 & 533 & 0.70 & 40 & 60 & 0.33 & 0.74 &
    &  1.83 & 4.85 &  1.51 & 0.08 &
    &  2.21 &  0.78 &  1.32 &  0.21
    \\ 
    &EuMnO$_3$&
    & 767 & 556 & 0.71 & 38 & 54 & 0.35 & 0.75 &
    &  1.85 & 4.95 &  1.47 & 0.09 &
    &  2.27 &  0.79 &  1.35 &  0.22
    \\ 
    &GdMnO$_3$&
    & 772 & 561 & 0.72 & 43 & 48 & 0.39 & 0.73 &
    &  1.86 & 4.99 &  1.54 & 0.06 &
    &  2.32 &  0.81 &  1.32 &  0.24
    \\ 
    \hline\hline
  \end{tabularx}
  \label{tab:tb}
\end{table*}

Regarding the hopping parameters [Fig.~\ref{fig:modelparhop} and
Tab.~\ref{tab:tb}], the nearest neighbor hoppings
$t^{\uparrow\uparrow}$ and $t^{\downarrow\downarrow}$ (calculated
using the purely JT distorted structure) exhibit a slight monotonic
increase with $R$ for PBE and HSE, which can be attributed to the unit
cell volume reduction. This trend is not followed for early $R$ series
members at \gowo. The deviation is not as pronounced for
$t^{\uparrow\uparrow}$ as it is for $t^{\downarrow\downarrow}$, but in
general, as mentioned above, results for the minority bands at \gowo\
should be taken with much care. The increase in $t^{\uparrow\uparrow}$
from the PBE values to those at beyond PBE levels is due to the
stronger hybridization with lower lying O $p$
states~\cite{2011-08//kovacik/ederer,2012-06//franchini/kovacik//kresse}. The
strong reduction of the hopping amplitude due to the increasing GFO
distortion along the $R$ series can be seen in the plots of
$t^{\uparrow\uparrow}$ and $t^{\downarrow\downarrow}$ (calculated
using the \textit{Pbnm} structure) and in the corresponding derived
reduction parameters $\eta_t^\uparrow$ and $\eta_t^\downarrow$. While
the reduction is strongest at PBE, generally followed by HSE and
\gowo\ in the case of majority spin, reversed behavior can be seen for
the minority spin. The decrease of the hoppings correlates with the
reduction of Mn-O-Mn in-plane angle $\phi_{ab}$ and the N\'eel
temperature. The further neighbor hopping parameters $t^{xy}$ and
$t^{2z}$ remain nearly unchanged along the $R$ series for PBE and
HSE. Not much significance should be given to the irregularities
observed for \gowo, since the notorious difficulty to properly
converge the minority bands can have a very pronounced effect on these
parameters. The $\widetilde{\lambda}$ parameter, controlling the JT
induced splitting in the hopping matrix, is largely independent on $R$
(except \mbox{$R=\text{La}$} for HSE) and its magnitude increases from
PBE through \gowo\ to HSE, resembling the behavior of the on-site
parameters $\lambda^\uparrow$ and $J_\mathrm{H}$ (see below).

The on-site TB parameters as a function of $R$ calculated for
different XC kernel treatment are presented in
Fig.~\ref{fig:modelparonsite} and Tab.~\ref{tab:tb}. For completeness,
the numerical values of the majority spin eigenvalue splitting
$\Delta\varepsilon^\uparrow$ and occupation matrix splitting
$\Delta{n}^\uparrow$ needed for $U_\mathrm{W}^{(\lambda)}$ evaluation
are also listed in Tab.~\ref{tab:tb}. The majority spin JT coupling
strength $\lambda^\uparrow$ and the Hund's rule coupling strength
$J_\mathrm{H}$ are almost constant along the $R$ series. They also
exhibit a mutually consistent qualitative increase at the HSE and
\gowo\ levels (compare with $\widetilde{\lambda}$ above), which is
reflected in the TB model by an increase in the band gap
($\lambda^\uparrow$) and the spin splitting ($J_\mathrm{H}$). The
magnitude of $\lambda^{\downarrow}$ is significantly smaller than that
of $\lambda^{\uparrow}$. This can be explained using the simple
argument of the weaker $d$-$p$ hybridization for the higher lying
minority bands~\cite{2010-06//kovacik/ederer}. The irregularities
along the $R$ series for \gowo\ are again caused by the quality of the
minority spin bands. The reduction parameter of the JT coupling
strength due to the GFO distortion $\eta^\lambda$ is weaker than the
corresponding hopping reduction parameters ($\eta^\uparrow$ and
$\eta^\downarrow$). Its relative change down the $R$ series is
comparable for PBE and HSE, while the results for \gowo\ are almost
$R$ independent.

In order to capture both the spin splitting and the band gap increase
when using HSE and \gowo, a single interaction parameter
$U_\mathrm{W}^{(\lambda)}$ is not sufficient and a semi-empirical
correction to the Hund's coupling $\Delta{J}_\mathrm{W}^{(\lambda)}$
is also needed in Model~2. Here, the hopping parameters of the
respective method are used, while the on-site parameters are kept
fixed at their PBE values. The el-el interaction parameter
$U_\mathrm{W}^{(\lambda)}$ as a function of $R$ can be regarded as a
constant for HSE, while it exhibits a small increase in the case of
\gowo. The value of $U_\mathrm{W}^{(\lambda)}$ is significantly
different for HSE and \gowo. The much larger HSE value can be a
consequence of the mixing parameter used in these
calculations~\cite{2012-12//he/franchini,2014-06//franchini}, which
consistently leads to an overestimation of the band gap (see
Table~\ref{tab:mm-bg}). The quantitatively less important
$\Delta{J}_\mathrm{W}^{(\lambda)}$ follows the same trend as
$U_\mathrm{W}^{(\lambda)}$.

\section{Conclusions}

A combination of first-principles calculations and tight-binding (TB)
model Hamiltonian via Maximally localized Wannier functions (MLWFs)
was applied to the parent compounds of manganites $R$MnO$_{3}$ ($R$ =
La, Pr, Nd, Sm, Eu and Gd). The electronic and magnetic properties
were studied at different levels of XC treatment.

The band structures within the same XC level exhibit similar features
along the $R$ series. The calculations show a clear trend of an
increase of the electronic band gap with the decrease of the $R$
cation radius $r_R$. While PBE band gaps are severely underestimated,
the HSE values are overestimated likely due to the amount of the exact
exchange included in the functional. The values obtained for \gowo\
seem to be more consistent with the available experimental
data. Likewise, the dielectric function calculated within \gowo\ is in
reasonable qualitative agreement with experiments but the intensity of
the first peak and $\epsilon_\infty$ are significantly overestimated.

The exchange couplings obtained at the HSE level yield Monte Carlo
simulated N\'eel temperatures that are in very good agreement with
experimental observations. The weakening of the FM in-plane exchange
interaction parameter with decreasing $r_R$ is a clear indication of
the destabilization of the A-type AFM order towards the E-type AFM
order observed in further members of the $R$ series. Concurrently, the
effective AFM out-of-plane exchange interaction strengthens and it is
only in LaMnO$_3$ where the out-of-plane antiferromagnetism can not be
attributed to a single exchange parameter.

Despite the difficulties in the disentanglement of the $e_{g}$
character states mainly at the \gowo\ level, the obtained MLWF bands
are in very good agreement with the underlying \emph{ab initio}
bands. The method-derived changes in the TB parameters due to
different treatments of the XC kernel has been investigated and
accounted for using two parameterization models. In general, an
overall consistent qualitative trend in the description of the TB
parameters has been found for all the compounds down the $R$ series at
the PBE, HSE and \gowo\ levels. The trends in the nearest neighbor
hopping amplitudes in the \textit{Pbnm} structure are comparable with
those of the volume, tolerance factor, Mn-O-Mn bond angles and the
N\'eel temperature. Another interesting result is that the JT and
Hund's rule coupling strength, as well as the simple mean-field
electron-electron interaction parameter, are practically $R$
independent and can be regarded as method dependent universal
constants in the $R$MnO$_{3}$ series.

\begin{acknowledgments}
  This work has been supported by the 7$^{\text{th}}$ Framework
  Programme of the European Community, within the project ATHENA. Part
  of the calculations were performed at the Vienna Scientific Cluster
  (VSC2).
\end{acknowledgments}

\bibliography{mybibfile}

\end{document}